\documentstyle[sprocl,epsfig]{article}

\def\be{\begin{equation}}
\def\ee{\end{equation}}
\def\bea{\begin{eqnarray}}
\def\eea{\end{eqnarray}}

\newcommand{\bs}{\bf}

\begin{document}

\title{A DYNAMICAL MODEL FOR THE RESONANT MULTIPOLES AND THE $\Delta$
 STRUCTURE }

\author{S. S. KAMALOV\footnote{Permanent address: Laboratory
of Theoretical Physics, JINR Dubna,141980 Moscow region, Russia} and
Shin Nan YANG}

\address{Department of Physics, National Taiwan University, Taipei
10617, Taiwan}

\maketitle

\abstracts{We show that recent experiment data for the ratios
$E_{1^+}/M_{1^+}$ and $S_{1^+}/M_{1^+}$  can be explained in a
dynamical model for electromagnetic production of pions, together
with a simple scaling assumption for the bare $\gamma^* N \Delta$
form factors. Within our model we find that the bare $\Delta$ is
almost spherical and the electric E2 and Coulomb C2 quadrupole
excitations of the physical $\Delta$ are nearly saturated by pion
cloud contribution in $Q^2 \le 4.0\,\,GeV^2$. }

\section{Introduction}

The study of excitations of the
hadrons can shed light on the nonperturbative aspects of QCD. One
case which has recently been under intensive study is the
electromagnetic excitation of the $\Delta(1232)$ resonance. At low
four-momentum transfer squared $Q^2$,  the interest is motivated
by the possibility of observing a $D$ state in the $\Delta$. The
existence of a $D$ state in the $\Delta$ has the consequence that
the $\Delta$ is deformed and the photon can excite a nucleon
through electric $E2$ and Coulomb $C2$ quadrupole transitions.
In a symmetric SU(6) quark model, the electromagnetic excitation of
the $\Delta$ could proceed only via magnetic $M1$ transition. In pion
electroproduction, $E2$ and $C2$ excitations would give rise to
nonvanishing $E_{1+}^{(3/2)}$ and $S_{1+}^{(3/2)}$ multipole
amplitudes. Recent experiments give nonvanishing ratio  $R_{EM} =
E_{1+}^{(3/2)}/M_{1+}^{(3/2)} \sim -0.03$~\cite{Beck97} at $Q^2=0$
which has been widely taken as an indication of the $\Delta$
deformation.

At sufficiently large $Q^2$, the perturbative QCD (pQCD) is
expected to work. It predicts that only helicity-conserving
amplitudes contribute at high $Q^2$, leading to
 $R_{EM} = E_{1+}^{(3/2)}/M_{1+}^{(3/2)} \rightarrow 1$ and
 $R_{SM} = S_{1+}^{(3/2)}/M_{1+}^{(3/2)} \rightarrow const$.
This behavior in the perturbative domain is very different from
that in the nonperturbative one. It is an intriguing question to
find the region of $Q^2$ which signals the onset of the pQCD.

In a recent measurement \cite{Frolov99}, the electromagnetic
excitation of the $\Delta$ was studied at $Q^2=2.8$ and $4.0
\,\,GeV^2$ via reaction $p(e,e'p)\pi^0$. The extracted ratios
$R_{EM}$ and $R_{SM}$ remain small and {\it negative}. This
disagrees with the previous analysis \cite{Burkert95} of the
earlier DESY data which gave small but {\it positive} $R_{EM}$ and
$R_{SM}$ at $Q^2 = 3.2 \,\,GeV^2$, though both analyses indicate
that pQCD is still not applicable in this region of $Q^2$. In this
talk, we want to show~\cite{Kamalov99} that the recent data of Ref. 2
can be understood from the dominance of the pion cloud
contribution at low $Q^2$ in both $E_{1+}^{(3/2)}$ and
$S_{1+}^{(3/2)}$ multipoles, as predicted by a dynamical model
\cite{Tanabe85,Yang85} for electromagnetic production of pion,
together with a simple scaling assumption for the bare $\gamma^*
N\Delta$ form factors.

\section{Dynamical Model for $\gamma^*N\rightarrow\pi N$}

The main feature of dynamical approach to pion photo- and
electro-production \cite{Tanabe85,Yang85} is that the unitarity is
built in by explicitly including the final state $\pi N$
interaction in the theory, namely, t-matrix is expressed as
\begin{eqnarray}
t_{\gamma\pi}(E)=v_{\gamma\pi}+v_{\gamma\pi}\,g_0(E)\,t_{\pi
N}(E)\,, \label{eq:tgamapi}
\end{eqnarray}
where $v_{\gamma\pi}$ is the transition potential operator for
$\gamma^*N \rightarrow \pi N$, and $t_{\pi N}$ and $g_0$ denote
the $\pi N$ t-matrix and free propagator, respectively, with $E
\equiv W$ the total energy in the CM frame.

Multipole decomposition of Eq. (\ref{eq:tgamapi}) gives the
physical amplitude in channel $\alpha$~\cite{Yang85}
\begin{eqnarray}
t_{\gamma\pi}^{(\alpha)}(q_E,k;&E&+i\epsilon)
=\exp{(i\delta^{(\alpha)})}\,\cos{\delta^{(\alpha)}} \nonumber\\
&\times&\left[v_{\gamma\pi}^{(\alpha)}(q_E,k) + P\int_0^{\infty}
dq' \frac{q'^2R_{\pi
N}^{(\alpha)}(q_E,q';E)\,v_{\gamma\pi}^{(\alpha)}(q',k)}{E-E_{\pi
N}(q')}\right], \label{eq:backgr}
\end{eqnarray}
where $\delta^{(\alpha)}$ and $R^{(\alpha)}$ are the $\pi N$
scattering phase shift and reaction matrix in channel
$\alpha$, respectively; $q_E$ is the pion on-shell momentum and
$k=|{\bf k}|$ is the photon momentum.

The multipole amplitude in Eq. (\ref{eq:backgr}) manifestly satisfies the
Watson theorem and shows that $\gamma\pi$ multipoles depend on the
half-off-shell behavior of  $\pi N$ interaction.
We remark that the use of K-matrix unitarization scheme as employed
in, e.g., Ref. 7 would amount to approximating Eq. (\ref{eq:backgr})
with
\begin{eqnarray}
t_{\gamma\pi}^{(\alpha)}(q_E,k;E+i\epsilon)
=\exp{(i\delta^{(\alpha)})}\,\cos{\delta^{(\alpha)}}
v_{\gamma\pi}^{(\alpha)}(q_E,k).
\label{eq:kmatrix}
\end{eqnarray}
The difference between Eqs. (\ref{eq:backgr}) and (\ref{eq:kmatrix}) lies in
the fact that only the on-shell rescatterings are included in the K-matrix
unitarization scheme.

In the resonant (3,3) channel in which $\Delta(1232)$ plays a dominant role,
the transition potential $v_{\gamma\pi}$
consists of two terms
\begin{eqnarray}
v_{\gamma\pi}(E)=v_{\gamma\pi}^B + v_{\gamma\pi}^{\Delta}(E)\,,
\label{eq:tranpot}
\end{eqnarray}
where $v_{\gamma\pi}^B$ is the background transition potential
which includes Born terms and vector mesons exchange
contributions, as described in Ref. 8. The second term
of Eq. (\ref{eq:tranpot}) corresponds to the contribution of bare
$\Delta$.

With Eq. (\ref{eq:tranpot}), we may decompose Eq. (\ref{eq:tgamapi}) in the
following way
\begin{eqnarray}
t_{\gamma\pi}(E)=t_{\gamma\pi}^B + t_{\gamma\pi}^{\Delta}(E)\,,
\label{eq:ddecomp}
\end{eqnarray}
where
\begin{eqnarray}
t_{\gamma\pi}^B(E)&=&v_{\gamma\pi}^B+v_{\gamma\pi}^B\,g_0(E)\,t_{\pi
N}(E),\label{eq:decompa}\\
t_{\gamma\pi}^\Delta(E)&=&v_{\gamma\pi}^\Delta+v_{\gamma\pi}^\Delta\,g_0(E)\,t_{\pi
N}(E)\,. \label{eq:decompb}
\end{eqnarray}
Here $t_{\gamma\pi}^B$ includes contributions from the nonresonant
background  and renormalization on the  vertex $\gamma^*N\Delta$
 due to $\pi N$ scattering.
The advantage of such a decomposition is that all the processes
which start with the electromagnetic excitation of the bare
$\Delta$  are summed up in $t_{\gamma\pi}^\Delta$.

Multipole decomposition of Eq. (\ref{eq:decompa}) takes the same
form as Eq. (\ref{eq:backgr}) and is used to calculate  the
multipole amplitudes $M_{1+}^B(W,Q^2), \,E_{1+}^B(W,Q^2)$ and
$S_{1+}^B(W,Q^2)$
  with $R_{\pi N}^{(\alpha)}(q_E,q';E)$ obtained from a
 meson exchange model \cite{Hung94} for $\pi N$ interaction.
 Note that to make the principal value integration in Eq. (\ref{eq:backgr})
 associated with
$v_{\gamma\pi}^B$ convergent, we introduce an off-shell dipole
form factor with  cut-off parameter $\Lambda$=440 MeV. The gauge
invariance, violated due to the  off-shell rescattering effects,
is restored by the substitution $J_{\mu}^B \rightarrow J_{\mu}^B -
k_{\mu} k\cdot J^B/k^2$, where $J_{\mu}^B$ is the electromagnetic
current corresponding to the background contribution
$v_{\gamma\pi}^B$.

\section{$\gamma^*N\leftrightarrow \Delta$ transition form factors}

Now let us consider the $\Delta$ resonance contribution $t_{\gamma\pi}^{\Delta}$
in Eq.
(\ref{eq:ddecomp}). In keeping with the standard way of
experimental analysis and constituent quark model (CQM) calculations, we describe the
resonant multipoles $t_{\gamma\pi}^{\Delta,\alpha}$ with a
Breit-Wigner type of energy dependence, as was done
in the isobar model of Ref. 8,
\begin{equation}
t^{\Delta,\alpha}_{\gamma\pi}(W,Q^2)\,=\,{\bar{\cal A}}^{\Delta}_{\alpha}(Q^2)\,
\frac{f_{\gamma \Delta}\,\Gamma_{\Delta}\,M_{\Delta}\,f_{\pi \Delta} }
 {M_{\Delta}^2-W^2-iM_{\Delta}\Gamma_{\Delta}}\,e^{i\phi}\,,
\label{eq:resonance}
\end{equation}
where $f_{\pi \Delta}(W)$ is the usual Breit-Wigner factor
describing the decay of the $\Delta$ resonance with total width
$\Gamma_{\Delta}(W)$ and physical mass $M_{\Delta}$=1232 MeV. The
expressions for $f_{\gamma \Delta}, \, f_{\pi \Delta}$ and
$\Gamma_{\Delta}$ are taken from Ref. 8. The phase
$\phi(W)$ in Eq. (\ref{eq:resonance}) is to adjust the phase of
$t_{\gamma\pi}^{\Delta,\alpha}$ to be equal to the corresponding pion-nucleon
scattering phase $\delta^{(33)}$. Note that at the resonance
energy $\phi(M_{\Delta})=0$.

\begin{figure}[tbp]
\begin{center}
\epsfig{file=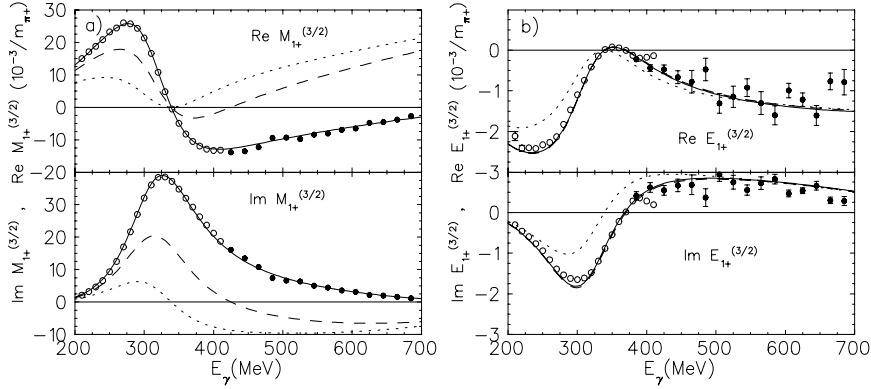,width=2in,height=4.5in, angle=90}
\end{center}
\vspace{-0.0cm} \caption{\small Real and imaginary parts of the,
{\bf (a)} $M_{1+}^{(3/2)}$, and {\bf (b)} $E_{1+}^{(3/2)}$
multipoles. The open and full circles are the results from the
Mainz dispersion relation analysis~\protect\cite{HDT} and from the VPI
analysis~\protect\cite{VPI97}, respectively. Notations for the
curves are given in the text.} \label{fig1_ky} \end{figure}

The main parameters in the bare $\gamma^* N \Delta$ vertex are the
${\bar{\cal A}}^{\Delta}(Q^2)$'s in Eq. (\ref{eq:resonance}). For the
magnetic dipole ${\bar{\cal M}}^{\Delta}$ and electric
quadrupole ${\bar{\cal E}}^{\Delta}$ transitions, they are
related to the conventional electromagnetic helicity amplitudes
$A^\Delta_{1/2}$ and $A^\Delta_{3/2}$ by
\begin{eqnarray}
{\bar{\cal M}^\Delta}(Q^2)=-\frac{1}{2}(A^\Delta_{1/2} + \sqrt{3}
A^\Delta_{3/2})\,,\qquad{\bar{\cal
E}^\Delta}(Q^2)=\frac{1}{2}(-A^\Delta_{1/2} + \frac{1}{\sqrt{3}}
A^\Delta_{3/2})\,.
\label{eq:helicity}
\end{eqnarray}
At the photon point $Q^2=0$,  the bare amplitudes ${\bar{\cal
M}}^\Delta (0)$ and ${\bar{\cal E}}^\Delta (0)$ of Eq.
(\ref{eq:resonance}) are determined from the  best fit to the
results of the recent analyses of Mainz~\cite{HDT} (open circles)
and VPI group~\cite{VPI97} (full circles), as shown in Fig. \ref{fig1_ky}. 
The dashed curves denote the contribution from $t_{\gamma\pi}^B$
only. The dotted curves correspond to  the K-matrix approximation to
$t_{\gamma\pi}^B$, namely, without the inclusion of principal value integral
term. Solid curves are the full results of our calculation with bare
$\Delta$ excitation.

The numerical values  for ${\bar{\cal M}}^\Delta$ and ${\bar{\cal
E}}^\Delta$ and the helicity amplitudes, at $Q^2 =0$, are given in
Table 1. Here we also give "dressed" values obtained using
K-matrix approximation for $t_{\gamma \pi}^B$. One notices that
the bare values determined above  for the helicity amplitudes
amount to only about $60\%$ of the corresponding dressed
values and are close to the predictions of  the CQM, as
 pointed out by Sato and Lee \cite{Sato96}. The
large reduction of the helicity amplitudes from the dressed to the
bares ones results from the fact that the principal value integral
part of  $t_{\gamma \pi}^B$, which represents the effects of the
off-shell pion rescattering, contributes approximately half of
the $M_{1+}$ as indicated by the dashed curves in Fig. \ref{fig1_ky}.
\begin{table}[htbp]
\caption {\small Comparison of the "bare" and "dressed" values for
the amplitudes ${\bar{\cal A}}^{\Delta},\, A_{1/2}^{\Delta}$ and
$A_{3/2}^{\Delta}$ (in $10^{-3}\,GeV^{-1/2}$).}
\vspace{0.2cm}
\renewcommand{\tabcolsep}{1.5pc} 
\begin{tabular}{|c|ccc|}
\hline
             Amplitudes            &  "bare"         & "dressed"    & PDG \\
 \hline ${\bar{\cal M}}^{\Delta}$  & $ 158\pm 2 $    & $289\pm 2 $  & $293\pm 8$ \\
        ${\bar{\cal E}}^{\Delta}$  & $ 0.4\pm 0.3 $  & $-7\pm 0.4 $ & $-4.5 \pm 4.2 $ \\
        $ A_{1/2}^{\Delta}$        & $ -80 \pm 2 $   & $ -134\pm 2$ & $-140\pm 5 $\\
        $ A_{3/2}^{\Delta}$        & $ -136 \pm 3 $  & $ -256\pm 2$ & $-258\pm 6$\\
\hline
\end{tabular}
\end{table}

 We now turn to the $Q^2$ evolution of the multipoles
 ${\bar{\cal M}}^\Delta(Q^2)$ and $\bar{\cal E}^\Delta(Q^2)$.
 In the present work, we parametrize the $Q^2$
dependence of the dominant ${\bar{\cal M}}^\Delta$ amplitude by
\begin{eqnarray}
{\bar{\cal M}}^\Delta(Q^2)={\bar{\cal M}}(0) \frac{\mid {\bs
k}\mid}{k_{\Delta}}\,(1+\beta Q^2)\, e^{-\gamma Q^2}\, G_D(Q^2)\,,
\label{eq:q2ansatz}
\end{eqnarray}
where $G_D$ is the nucleon dipole form factor.  For the small
${\bar{\cal E}}^\Delta$ and ${\bar{\cal S}}^\Delta$ amplitudes,
 we follow Ref. 8 and assume that they have the same
$Q^2$ dependence as ${\bar{\cal M}}^\Delta$ ({\it scaling assumption}).
This is motivated by the scaling law which has been observed for the
nucleon form factors.

We remind the reader that, in contrast to Ref. 8, amplitudes
${\bar{\cal M}}_{1+}$ and ${\bar{\cal E}}_{1+}$ and the corresponding
helicity amplitudes in Eq. (\ref{eq:helicity}) correspond to the
"bare" $\gamma N \Delta$ transition.  For the real photon, they are
equal to the standard $M1$ and $E2$ amplitudes of the $\Delta
\leftrightarrow N\gamma$ transition defined in accordance with the
convention of the Particle Data Groups. At the resonance energy, they can
be easily expressed in terms of the Dirac-type form factors $g_1$
and $g_2$ used in Ref. 7, or Sachs-type form
factors $G_M$ and $G_E$ used in Ref. 12. The relation
between ${\bar{\cal M}}_{1+}$ amplitude and the bare $G_M$ form
factors is as follows
\begin{equation}
{\bar{\cal M}}_{1+}(Q^2)=\frac{e}{2m}\,\frac{\mid {\bf k} \mid}{k_{\Delta}}
\,\sqrt{\frac{k_{\Delta} M_{\Delta}}{m}}\,G_M (Q^2), \label{eq:gmform}
\end{equation}
where $k_{\Delta}=(M^2_{\Delta}-m^2)/2M_{\Delta}$ with $m$ and $M_{\Delta}$
denoting the nucleon and $\Delta$ mass, respectively. Expression for
the electric amplitude is similar, but with opposite sign.
Relation between physical $M_{1+}^{(3/2)}$ multipole and
experimentally measured $G_M^*$ form factor is given by Eq. (24)
of Ref. 8. Note that we employ the
"Ash" definition~\cite{Ash67} for the $G_M^*$ which differs from the
$\Delta$ form factor used in Ref. 2 by a factor
$f=\sqrt{1+Q^2/(m+M_{\Delta})^2}$, i.e.,
$G_M^*(our)=G_M^*($Ref. 2)$/f$.

\section{Results and Discussion}

\begin{figure}[htb]
\begin{center}
\epsfig{file=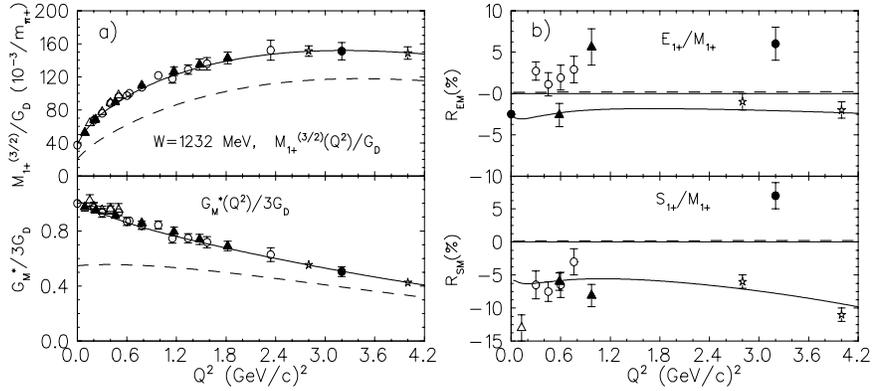,width=2.in,height=4.5in, angle=90}
\end{center}
\vspace{-0.0cm}\caption{\small The $Q^2$ dependence of {\bf (a)}
$Im\,M_{1+}^{(3/2)}$ with corresponding $G_M^*$ form factor, and
{\bf (b)} ratios $E_{1+}^{(3/2)}/M_{1+}^{(3/2)}$ and
$S_{1+}^{(3/2)}/M_{1+}^{(3/2)}$  at $W$=1232 MeV. The full and
dashed curves are the results for the "dressed" and "bare" $\gamma
N \Delta$ vertexes, respectively. Experimental data as quoted in
Ref. 8. The new data at $Q^2$=2.8 and 4.0
$(GeV/c)^2$ are from Ref. 2.} \label{fig2_ky}
\end{figure}

Using the  $\beta$ and $\gamma$ in Eq. (\ref{eq:q2ansatz}) as free
parameters, we fit the recent experimental data \cite{Frolov99} as
well as old one quoted in Ref. 8 on the $Q^2$ dependence
of the $M_{1+}^{(3/2)}$ or equivalently, the $G_M^*$
form factor. Our result is shown in Fig. \ref{fig2_ky}. The
obtained values for the $\beta$ and $\gamma$ parameters are:
 $\beta=0.44 \,\,GeV^{-2}$ and  $\gamma=0.38\,\, GeV^{-2}$. Here the
dashed curve corresponds to contribution from the  bare $\Delta$,
i.e., $t_{\gamma\pi}^\Delta$ of Eq. (\ref{eq:decompb}).
The results for the ratios $R_{EM}=E_{1+}^{(3/2)}/M_{1+}^{(3/2)}$
and $R_{SM}=S_{1+}^{(3/2)}/M_{1+}^{(3/2)}$ are  shown in the right
column of Fig. \ref{fig2_ky}.
It is seen that they are in good agreement with the results of the
model independent analysis of Ref. 2 up to $Q^2$ as
high as $4.0\,\,GeV^2$. Note that since the bare values for the
E2 and C2 excitations are small, the absolute values
and shape of these ratios are determined, to a large extent, by
the pion rescattering contribution. The bare $\Delta$ excitation
contributes mostly to the $M_{1+}^{(3/2)}$ multipole.

 Pion cloud has been found to play an important role in hadron structure
 in many studies. For example, in the cloudy bag model (CBM), a
reasonably good agreement~\cite{Kalberman83,Bermuth88} with the
measured $R_{EM}$ can be obtained with bag radius of $R= 0.6 -0.8\,\,
fm.$ In a recent  improved CBM calculation~\cite{Lu97}, where the
relativistic effects and CM motion were better treated, it was
also found that the pion cloud gives large contribution to
$G_M^*(Q^2)$. It is generally concluded that the smaller the bag
radius, the larger the pion cloud contribution. Similar conclusion
is reached with respect to the proton EM form factors within the
cloudy bag model. In fact, it has been found~\cite{Lu00} that the
recent Jlab data~\cite{Jones00} on the scaling violation in the
electromagnetic form factors can be explained within CBM with $R=
0.7\,\,fm.$ One might then be tempted to interpret our result as another
indication of  preferring a
smaller bag radius. However, at these small bag radii, the pion field is
so strong that the use of the perturbative approach employed in
these CBM studies is questionable. 
 In this connection, we should
mention that in a nonperturbative calculation  within a chiral
chromodielectric model and a linear $\sigma$-model, Fiolhais,
Golli, and Sirca~\cite{Fiolhais96} reached a similar conclusion
as ours, namely, the large
experimental values of $R_{EM}$ and $R_{SM}$ could be explained in
terms of the pion contribution alone.

In nonrelativistic CQM it is well known that, the values of $R_{EM}$ and
$R_{SM}$ obtained with a $D-$state admixture in the $\Delta$
generated by one-gluon-exchange hyperfine interaction are in general too small.
It has recently been suggested by the T\"ubigen group~\cite{Buchman98} that
this problem can be fixed with the inclusion of exchange currents in the
calculation. The reason is that two-body exchange currents could flip the spins
of the two quarks in the nucleon to convert it into a $\Delta$. Such a
transition would just be a transition between the S-states in the nucleon and
$\Delta$, and accordingly  is greatly enhanced. We would like to point out here that the
diagram involving a two-body exchange current induced by the exchange of a pion
between two quarks in the nucleon can also be interpreted as  pion rescattering
effects, as considered in the current study.

\section{Summary}

In summary, we calculate the $Q^2$ dependence of the ratios
$E_{1+}/M_{1+}$ and $S_{1+}/M_{1+}$ in the
$\gamma^*N\rightarrow\Delta$ transition, with the use of a
dynamical model for electromagnetic production of pions. We find~\cite{Kamalov99}
that both ratios $E_{1+}/M_{1+}$ and $S_{1+}/M_{1+}$ remain small
and negative for $Q^2 \le 4.0\,\,GeV^2$. Our results agree well
with the recent measurement of Frolov et al. \cite{Frolov99}, but
deviate strongly from the predictions of pQCD. Our results
indicate that the bare $\Delta$ is almost spherical and hence very
difficult to be directly excited via electric E2 and Coulomb C2
quadrupole excitations. The experimentally observed
$E_{1+}^{(3/2)}$ and $S_{1+}^{(3/2)}$ multipoles are, to a very
large extent, saturated by the contribution from pion cloud, i.e.,
pion rescattering effects. It remains an intriguing question, both
experimentally and theoretically, to find the region of $Q^2$
which will signal the onset of pQCD.

\section*{Acknowledgments}
This work is supported in part by the NSC/ROC under the grant no.
NSC 88-2112-M002-015.



\section*{References}
\begin{thebibliography}{99}
\bibitem{Beck97} R. Beck et al., {\it Phys. Rev. Lett.} {\bf 78}, 606 (1997);
G. Blanpied et al. {\it Phys. Rev. Lett.} {\bf 79}, 4337 (1997).
\bibitem{Frolov99} V.V. Frolov et al., {\it Phys. Rev. Lett.} {\bf 82},
45 (1999).
\bibitem{Burkert95} V. Burkert and L. Elouadrhiri, {\it Phys. Rev. Lett.} {\bf 75},
 3614 (1995).
\bibitem{Kamalov99}S.S. Kamalov and S.N. Yang, {\it Phys. Rev. Lett.} {\bf 83},
 4494 (1999).
\bibitem{Tanabe85} H. Tanabe and K. Ohta, {\it Phys. Rev.} C {\bf 31}, 1876
(1985).
\bibitem{Yang85} S.N. Yang, {\it J. Phys.} G {\bf 11}, L205 (1985).
\bibitem{Davidson91}R. Davidson, N. C. Mukhopadhyay and R. Wittman,
Phys. Rev.  {\bf D43}, 71 (1991).
\bibitem{UIM} D. Drechsel, O. Hanstein, S.S. Kamalov and L. Tiator,
{\it Nucl. Phys.} A {\bf 645}, 145 (1999).
\bibitem{Hung94} C.T. Hung, S.N. Yang and T.-S.H. Lee, {\it J. Phys.} G
{\bf 20}, 1531 (1994); C. Lee, S.N. Yang and T.-S.H. Lee, {\it
ibid.} G {\bf 17}, L131 (1991).
\bibitem{HDT} O. Hanstein, D. Drechsel, and L. Tiator,
{\it Nucl. Phys.} A {\bf 632}, 561 (1998).
\bibitem{VPI97} R.A. Arndt, I.I. Strakovsky and R.L. Workman,
 {\it Phys. Rev.} C {\bf 53}, 430 (1996) (SP97 solution of the VPI analysis).
\bibitem{Sato96} T. Sato and T.-S.H. Lee, {\it Phys. Rev.} C {\bf 54}, 2660 (1996);
T.-S. H. Lee, in $N^*$ {\it Physics}, eds. T.-S.H. Lee and W.
Roberts, p. 19, (World Scientific, Singapore 1997).
\bibitem{Davidson86} R. Davidson, N. C. Mukhopadhyay and R. Wittman,
{\it Phys. Rev. Lett.} {\bf 56}, 804 (1986).
\bibitem{Ash67}  W. W. Ash, {\it Phys.Lett.} {\bf 24B}, 165 (1967).
\bibitem{Kalberman83}G. K\"albermann and J.M. Eisenberg, {\it Phys. Rev.} D {\bf 28},71
 (1983).
\bibitem{Bermuth88}K. Bermuth, D. Drechsel, L. Tiator, and J.B. Seaborn,
{\it Phys. Rev.}  D {\bf 37}, 89 (1988).
\bibitem{Lu97}D.H. Lu, A.W. Thomas, and A.G. Thomas, {\it Phys. Rev.} C {\bf 55},
3108  (1997).
\bibitem{Lu00}D.H. Lu, S.N. Yang, and A.W. Thomas, {\it J. Phys.} G (2000),
 in print.
\bibitem{Jones00}M. K. Jones et al, {\it Phys. Rev. Lett.} {\bf 84} 1398 (2000).
\bibitem{Fiolhais96}M. Fiolhais, B. Golli, and S. Sirca, {\it Phys. Lett.} B
{\bf 373} 229 (2000).
\bibitem{Buchman98}A.J. Buchmann, E. Hern\'andez, U. Meyer, and A. Faessler,
{\it Phys. Rev.} C {\bf 58}, 2478  (1998);  A.J. Buchmann, "Baryons' 98", eds.
D. W. Menze and B. Ch. Metsch, World Scientific, Singapore, 1999, p. 731
(hep-ph/9909385); and this proceeding.
\end{thebibliography}
\end{document}